\documentclass[12pt,preprint]{aastex}

\shorttitle{}
\shortauthors{N. Okabe and M. Hattori}

\begin{document}

\title{ A SPONTANEOUS GENERATION OF THE MAGNETIC FIELD AND SUPPRESSION
OF THE HEAT CONDUCTION IN COLD FRONTS}

\author{Nobuhiro OKABE and Makoto HATTORI}
\affil{Astronomical Institute, Tohoku University, Sendai 980-8578, Japan}
\email{okabe@astr.tohoku.ac.jp}
\email{hattori@astr.tohoku.ac.jp}

\begin{abstract}

We have determined the physical mechanism responsible for the plasma instabilities,
which was first found by Ramani and Laval (1978), 
associated with anisotropic velocity distributions induced by the
temperature gradient in which
there are growing low frequency transverse magnetic 
waves, even in the absence of  background magnetic fields.
We have shown that the physical mechanism responsible for the growth of one of the 
modes is 
identical to the Weibel instability. 
The nonlinear saturation level of the
 instability is also provided by considering the wave-particle interactions. 
The non-linear evolutions of the magnetic fields after the saturation are 
speculated. 
The results are applied to the cold fronts which is 
 one of the newly discovered structures in clusters of galaxies by the Chandra X-ray observatory. 
We predict the
 existence of the magnetic field of  $\sim 10\mu$G tangential to the surface over the entire region
of the cold front surface 
and that the   
heat conduction is significantly suppressed by the trapping of the electrons by the 
generated magnetic fields.   
The instability may provide a new possibility on the origin of cosmic magnetic field.
\end{abstract}


\keywords{galaxies: clusters: general---magnetic
fields---conduction---instabilities---plasmas}


\section{INTRODUCTION}

The Chandra X-ray observatory revealed the bow-shaped
discontinuities in the X-ray emitting hot plasmas in the clusters of
galaxies \cite{mar00}. The plasma temperature sharply increases across the discontinuities, while 
the plasma density sharply decreases across the discontinuities toward
the same direction. These new structures  
are called cold fronts.  The cold fronts  
might be  the contacting surfaces of two different origin plasmas. If the
classical Spitzer heat conductivity 
is applied, the life time of the
cold fronts should be $\sim 10^6$yr.  Since it is much less than the
expected ages of the cold fronts, that is $>10^8$yr, the heat
conductivity in the cold fronts region must be significantly reduced
from the classical Spitzer value \cite{ett00,mar00}. Since there are
thermal pressure jumps across the fronts (Vikhlinin, Markevitch \& Murray 2001a),  a transonic motion of the
cold fronts is expected to maintain the pressure balance between the
cold and hot regions. Vikhlinin et al. (2001b, 2002) pointed out that 
the magnetic field of $10\mu$G along the fronts should exist
to explain the smoothness of the fronts otherwise the irregular structures are
expected due to the Kelvin-Helmholtz (KH) instability. Since the
required field strength is more than order of magnitude stronger than
the intracluster magnetic field strength while the plasma density jump
is at most factor of two across the cold fronts, the compression of the
intracluster magnetic field as conserving the flux can not explain the
existence of such a strong magnetic field in the cold fronts.

A precise mathematical treatment of the thin hot plasmas based on 
the plasma kinetic theory found that the anisotropic electron  
velocity distribution induced by the temperature gradient 
drives the plasma
instability,  and it is postulated that the electron scattering by the excited plasma waves
may significantly reduce the heat conductivity (Ramani \& Laval 1978).
The low frequency transverse growing magnetic
waves are excited even in the absence of an background magnetic field. 
Since the instability was found by Ramani and
Laval (1978), the instability is referred to the Ramani-Laval (RL) instability in this paper.
The applications of the RL instability to 
the cluster hot plasmas were examined by some authors; in cases of the
plasma with background magnetic fields \cite{lev92} and the electron-ion
two components plasma \cite{hat00}. 
Although the RL instability was said to be the Weibel like 
instability \cite{wei59} since the growth of the waves is seeded by the 
anisotropic velocity distribution, 
there are plenty of the qualitative differences 
among the both instabilities as explained in section 3. 
Therefore, the both instabilities were distinguished \cite{gal91} and
it is very important to determine the physical mechanism of the RL instability. 

Since this paper is the first paper of our series of papers,
we focus on the identification of the physical
mechanism of the RL instability (in \S 2. and 4.) including the review of the Weibel
instability (in Appendix) and the RL instability (in \S 3),
the determination of the non-linear saturation
level of the excited waves 
and the non-linear evolution of the excited modes (in \S 5), 
and the application to the cold fronts (in \S 6.). Section 7 is devoted to discussion. 

\section{THE VELOCITY DISTRIBUTION FUNCTION IN THE PLASMA WITH THE HEAT FLUX}

In this section, how the anisotropic electron velocity distribution function is 
set up when the temperature inhomogeneity exists in the plasma, 
is discussed from physical point of view. The absence of a background magnetic field is assumed. 
Consider the temperature inhomogeneity in hot electron plasma with  the temperature
variation scale of $L$.  
Since the heat conduction carries the heat flux from the hotter to cooler 
regions, the heat flux along the temperature gradient takes a negative 
finite value; $q  \propto \langle v_{\parallel} v^2 f\rangle < 0$ where the bracket denotes the
integral over the velocity space and  the subscript $\parallel$ denotes the component parallel to
 the temperature gradient.  Therefore, the electron velocity distribution function must deviate
from the Maxwell-Boltzmann $f_m=n_0(x_\parallel)(\pi
 v_{\rm th}(x_\parallel)^2)^{-3/2} \exp(-v^2/v_{\rm
 th}(x_\parallel)^2)$; $\Delta f =f- f_m \neq 0$. 
 Here, $n_0(x_\parallel)$ is the electron number density, $v_{\rm
 th}(x_\parallel)=(2 k_B T(x_\parallel)/m_e)^{1/2}$ is the thermal
 velocity with the temperature $T(x_\parallel)$ and the electron mass
 $m_e$.  The above condition
 together with $\langle \Delta f\rangle=0$, that is the number density
conservation,  and $\langle v^2 \Delta
 f\rangle=0$, that is the energy conservation,  restrict the form
 of the deviated part,  $\Delta f$, to be odd function of  the velocity component
 along the temperature gradient, $v_\parallel$. Of course, $\Delta
 f\rightarrow 0 $ as $|v_\parallel| \rightarrow \infty$. In Figure 1,
 the possible cases as $\Delta f$ are shown. From zero current condition,
 such as $\langle v_\parallel \Delta f\rangle=0$, the form like the type
 A is rejected  since $\Delta f$ where $v_\parallel>0$
 and $v_\parallel <0$ are respectively positive  and negative, and then
 $v_\parallel \Delta f>0$ for all velocity and $\langle v_\parallel \Delta f  \rangle >0$.  On the other
 hand, the type B and C satisfy the zero current condition since
 $\Delta f$ curve acrosses the $v_\parallel$-axis in each positive and negative $v_\parallel$ region. 
The $\langle v_{\parallel}
 v^2 f \rangle $ for the type B and C have finite values because this is the weighted mean 
of $\langle v_\parallel \Delta f  \rangle $ weighted by  $v^2$ which is larger when $|v_\parallel|$ is larger.
The heat flux condition of $\langle v_{\parallel} v^2 f \rangle < 0$
says that $\Delta f$ where $v_\parallel \rightarrow -\infty$ should be positive. 
Thus, the type C is  only the possible form as the deviated part in the plasma where the finite heat flow
exists.
The relative amplitude of the deviated part to the Maxwell-Boltzmann should be 
$\epsilon \delta_T$ where $\epsilon=\lambda_{\rm mfp}/L$ and $\delta_T=\delta T/T$ is the fractional temperature fluctuation, 
since the deviation may be  induced by the Coulomb collision and the temperature fluctuation.

The deviated part $\Delta f$ can be also deduced analytically.
The Boltzmann equation is 
\begin{eqnarray}
 \frac{\partial f}{\partial t}+v_\parallel \frac{\partial f}{\partial x_\parallel} 
- \frac{e}{m}E_\parallel \frac{\partial f}{\partial v_\parallel}=-\nu (f-f_{m}), \nonumber 
\end{eqnarray}
where $E_\parallel$ is the zeroth electric field along the temperature
gradient and  the rhs is the
Krook operator as the collision term and  $\nu=(k_B T/m_e)^{1/2}/\lambda_e$ is
the Coulomb collision frequency with the Coulomb mean free path
$\lambda_e$ \cite{sar88}. Hereafter, we describe the collision
frequency as $\nu=v_{\rm th}/ \lambda_{\rm mfp}$ with the parameters $v_{\rm th}=(2 k_B
T/m_e)^{1/2}$ and $\lambda_{\rm mfp}=\sqrt{2} \lambda_e$. For simplicity, the pressure balance is assumed.  Then,
$E_{\parallel}=0$ \cite{ram78}. If the perturbative treatment is applicable to describe the system, 
 the distribution function could be expanded in 
  $\epsilon\delta_T$ \cite{cha60} as,
\begin{eqnarray}
 f=f_m+\epsilon \delta_T f^{(1)}+ \epsilon^2 \delta_T^2 f^{(2)} \dots, \nonumber 
\end{eqnarray}
where $f^{(j)} \ (j=1,2\cdots)$ describe the deviation of the
distribution function from the Maxwell-Boltzmann in order of $(\epsilon \delta_T)^j$.
This expansion is known as the Chapman-Enskog expansion.
Therefore,  the electron distribution function up to the first order in $\epsilon\delta_T$
is obtained as 
\begin{eqnarray}
 f= f_m\left[1+\epsilon\delta_T\frac{v_\parallel}{v_{\rm th}}\left( \frac{5}{2}-
 \frac{v^2}{v_{\rm th}^2}\right)\right]. \nonumber 
\end{eqnarray}
The form of the deviated part is essentially the same as the type C shown
in Figure 1. We would like to note that adopting the Krook operator is not essential for determining the form
of the deviated part.

\section{THE REVIEW OF THE RL INSTABILITY}

The diagnostics of the Ramani-Laval (RL) instability are summarized contrasting
with the Weibel instability.  
As summarized in the Appendix,  the temperature anisotropy 
excites the transverse magnetic waves due to the Weibel instability. 
The excited wave is standing wave with zero phase velocity.
The amplitude of the 
wave grows when the temperature perpendicular to the wave vector is higher 
than the temperature parallel to the wave vector. 
The amplitude of the wave is damped in the opposite situation.  

Ramani and Laval (1978) studied the stability of the plasmas when the 
temperature distribution is not homogeneous. 
The non-equilibrium electron velocity distribution function 
is deduced from the assumed temperature distribution self consistently  
using the Chapman-Enskog expansion. 
This is one of the prominent difference from the Weibel 
case where the anisotropic electron velocity distribution 
function due to the temperature anisotropy is given by hand.
The difference of the deduced non-equilibrium velocity 
distribution function from the equilibrium one, that is the Maxwell-Boltzmann
distribution, is skewed in the direction of the temperature gradient
and is the odd function of the velocity component along the temperature 
gradient.  The velocity
distribution functions in the both cases are anisotropic. 
However, the skewed nature 
found in the RL case is not found in the Weibel.  
Although the anisotropic velocity dispersion is essential for the Weibel instability,
the velocity dispersion is isotropic in the RL case. 
Ramani and Laval (1978) performed the linear stability analysis
of the plasmas with the deduced non-equilibrium velocity
distribution following the procedure of the plasma kinetic theory. 
Two independent modes appear. The two modes are distinguished
whether the magnetic field is in  or not in  the 
plane made by the temperature gradient and the wave vector,
where the first one is named mode 2 and the second one is named mode 1.
The dispersion relation of the both modes have a real part  as
\begin{eqnarray*}
\omega_r&=&{\epsilon\delta_T\over 4}kv_{\rm th}{\rm cos}\theta.
\end{eqnarray*}
The imaginary part of the mode 1 is obtained as,
\begin{eqnarray*}
\gamma&=&{\epsilon^2\delta_T^2\over 8\sqrt{\pi}}kv_{\rm th}(3{\rm cos}^2\theta
-2{\rm sin}^2\theta)-{1\over \sqrt{\pi}}\left({c\over \omega_{\rm p}}\right)^2
k^3v_{\rm th},
\end{eqnarray*}
and  the imaginary part of the mode 2 is obtained as,
\begin{eqnarray*}
\gamma&=&{3\epsilon^2\delta_T^2\over 8\sqrt{\pi}}kv_{\rm th}{\rm cos}^2\theta
-{1\over \sqrt{\pi}}\left({c\over \omega_{\rm p}}\right)^2
k^3v_{\rm th},
\end{eqnarray*}
where $\omega_{\rm p}$ is the electron plasma frequency, $c$ is the speed of light  and
$\theta$ is the angle between the direction of the temperature 
gradient and the wave vector.
The existence of the real part in the dispersion relation of the RL instability 
is one of the   difference from the Weibel case.   
Since the phase velocity along the temperature gradient  
$\omega_r/k{\rm cos}\theta={\epsilon\delta_T\over 4}v_{\rm th}$ only takes a positive value,
the wave propagates only one way from the low temperature region to the high temperature
region.  
Therefore, the excited waves do not carry the heat from the hot to the cold region.
Although the mode 2 is a pure transverse wave,  the longitudinal component
of the electric fields, $E_k$, has a non-zero value for the mode 1 as
\begin{eqnarray*}
E_k&\sim&{\epsilon\delta_T\over 4}{\rm sin}\theta {v_{\rm th}\over c}B_z.
\end{eqnarray*}
This is an another prominent difference from the Weibel case.
The wave grows the most rapidly when the wave vector is 
parallel to the temperature gradient. 
The wave is damped when the direction of the wave propagation  is 
perpendicular to the temperature gradient. 
The dependence on the $k$ of the growth rate is the same as that of the Weibel. 

In a sence that the anisotropic velocity distribution is the driving force of the 
instability, the both instabilities are similar and therefore the RL instability was 
said to be the Weibel like.  However, there are plenty of the qualitative differences 
among the both instabilities as explained in above. 
Therefore, the both instabilities were distinguished \cite{gal91} and 
we try to identify the physical mechanism of the 
RL instability in the next section.

\section{THE PHYSICAL MECHANISM OF THE RL INSTABILITY}

The physical mechanism of the RL instability can be understood from the nature of 
the velocity distribution function when the finite heat flow exists, as illustrated in Figure 2.
In the following discussion, strictness of the numerical factor is ignored.
The peak position is shifted toward positive $v_{\parallel}$ direction and
the amount of the shift is $v_{\parallel}\sim \epsilon\delta_Tv_{{\rm th}}$.  
The phase velocity of 
any low frequency magnetic transverse wave must be 
close to the velocity at where the distribution function has a peak value
otherwise a finite net electric current is induced by the wave magnetic fields. 
This  explains why the excited waves by the RL instability have the phase 
velocity of $\epsilon\delta_Tv_{{\rm th}}$, and travel 
one way direction rising up the temperature gradient and  
the waves can not travel in the direction of the perpendicular to the 
temperature gradient. 
The reason why the excited waves by the Weibel instability  are standing wave,
is simply because there is no shift in peak position of the velocity 
distribution function in the Weibel case.
The peak value is increased by  $\sim 1+(\epsilon\delta_T)^2$  from  
that of the pure Maxwell-Boltzmann case since the amplitude of the 
deviated part takes a value of $\epsilon\delta_T v_{\parallel}\sim
(\epsilon\delta_T)^2v_{\rm th}$
relative to the Maxwell-Boltzmann and the value of the Maxwell-Boltzmann part is nearly same as the peak value
at $v_{\parallel}\sim \epsilon\delta_Tv_{{\rm th}}\ll v_{{\rm th}}$. 
Since the total electron number density must be unchanged, for the observer comoving 
with the waves this can be interpreted as 
the decrease of the effective electron temperature 
in the direction of the temperature 
gradient, $T_{\parallel}$,  by $\sim 1-(\epsilon\delta_T)^2$ 
relative to the temperature perpendicular to the 
temperature gradient, $T_{\perp}$ (Fig.2). 
The growth of the magnetic waves in the RL instability is, therefore, due to essentially the same mechanism as the 
Weibel instability \cite{wei59,fri59} in which the temperature anisotropy is the driving force 
of the instability. 
Consider the waves  traveling nearly parallel to the temperature gradient.
In this case, $T_{\perp,\vec{k}}\sim T_{\perp}$ and $T_{\parallel,\vec{k}}\sim T_{\parallel}$
where $T_{\perp,\vec{k}}$ and $T_{\parallel,\vec{k}}$ are the temperature components
perpendicular and parallel to the wave vector  
for the observers comoving with the waves,
respectively. 
Since $T_{\perp,\vec{k}} > T_{\parallel,\vec{k}}$, the waves can grow.  
As a result, the direction of the 
magnetic field generated by the instability is almost perpendicular
to the temperature gradient. 
The growth rate of the mode which travels 
toward the direction of the temperature gradient  with wavenumber $k$ is obtained from that of the Weibel
 instability \cite{kra73},
\begin{eqnarray}
\gamma&\sim& v_{{\rm th}}\left[\left({T_{\perp}\over T_{\parallel}}-1\right) k-\left({ck\over 
\omega_{{\rm p}}}\right)^2k\right]\nonumber\\
&\sim& v_{{\rm th}}\left[(\epsilon\delta_T)^2 k -\left({ck\over \omega_{{\rm p}}}\right)^2k\right]. \nonumber
\end{eqnarray}
The growth rate gets the maximum value of $\gamma_{\rm max}\sim (\epsilon\delta_T)^3 (v_{\rm th}/c)
\omega_{\rm p}$ at $k=k_{\rm max}\sim \epsilon\delta_T \omega_{\rm p}/c$.
When the direction of the wave vector is perpendicular to the temperature gradient,
$T_{\perp,\vec{k}}= T_{\parallel}$ and $T_{\parallel,\vec{k}}= T_{\perp}$.
Since $T_{\perp,\vec{k}} < T_{\parallel,\vec{k}}$ in this case, the wave can not grow. 
These results are exactly the same as the results led by the plasma kinetic theory
except numerical factors \citep{ram78, hat00, oka03}. 

The above discussion gives a complete physical explanation for the mode 2 of the RL instability.
However, these are not satisfactory to the mode 1 in which the longitudinal electric field component
appears. Unfortunately, we have not yet gotten the physical explanation for this.
This is still an open question.  

\section{ON THE NONLINEAR EVOLUTIONS OF THE RL INSTABILITY}

The nonlinear saturation level of the excited wave is estimated  assuming 
that the wave-particle interaction determines the saturation level.
The fundamentals are illustrated in Figure 3 \citep{ram78, gal91}.
Once the Larmor radius of an electron gets shorter than the wavelength
of the growing mode, the electron is trapped by the magnetic field of the wave
and the magnetic flux enclosed by its orbit
becomes finite. 
Then, the kinetic energy of the trapped electron 
starts monotonically increase as the growth of the magnetic field
strength, since the increase of the magnetic flux enclosed by the electron
orbit cause the induction electric fields which accelerate the electron as the
betatron accelerator. 
Once the Larmor radius of 
typical thermal electrons, $r_L\sim v_{\rm th}\omega_{\rm c}^{-1}$,
gets shorter than the wavelength of the fastest growing mode, that is 
$r_Lk_{\rm max}<1$, 
the increase of the kinetic energy of the electron system becomes significant 
if the waves still continue to grow. 
Since this finally violates the energy conservation, the growth of the magnetic field strength must be 
saturated when $r_Lk_{\rm max}\sim 1$. 

The evolution of the magnetic fields after the nonlinear saturation could be described as follows. 
Some numerical simulations which follow the evolution of the Weibel instability, showed that the
strength of the magnetic field driven by the Weibel instability decreases after it gets the maximum 
value \cite{mor71}.  
This can be understood as follows. 
As the magnetic field grows,
the electron velocity distribution is  isotropitized and it becomes difficult to 
maintain the electric current field which supports the magnetic field of the waves.  
In these simulations, the system is assumed to be isolated and 
the initial
anisotropic velocity distribution function let be free to evolve 
to isotropic one. 
On the other hand, 
in the case of the RL instability there is a driving force which  maintains 
the anisotropy of the velocity distribution
function. 
As far as the temperature gradient is not disappeared,  the finite heat flux transports the heat
from the  hot to the cold region and the anisotropic velocity distribution function
discussed in section 2 is maintained.  
Therefore, the decrease of the magnetic field strength after the nonlinear saturation as found in the 
Weibel instability may not be occurred in the RL instability. 
It is expected that the generated magnetic field strength is kept 
to be the saturated value for the life time of the temperature gradient.
Wallace and Epperlein (1991) performed  the numerical simulations to follow the evolution of the Weibel
instability when an initial anisotropic distribution function is maintained by an external
source.  They showed that the magnetic field strength is kept constant value 
after the saturation.  Their results support the above expectation for the RL instability.
There are several indicative numerical simulations concerning the organization of the globally
connected magnetic fields from the wavy fields generated by the instability.
The wavy magnetic 
fields generated by the Weibel instability 
evolve into the longer wavelength modes after the saturation \cite{lee73,sen00,sen02}.
This result indicates that the excited wavy magnetic field 
automatically evolves into the globally connected 
fields. 
This could be understood as follows.
After the magnetic field strength gets the saturated level, 
the electric current field starts to play as an individual electric beams every 
half wave length (Fig.4).  Each beam is surrounded by the azimuthal magnetic field 
generated by the current beam itself. 
The electric  beams  
interact  each other via the Amp\`ere's force \cite{sen00,sen02}.  
The beams directed to the same direction are attracted each other and automatically 
gather.  Finally, they merge into larger one beam. 
Since the physical mechanism of the growth of the RL instability
is the same as the Weibel instability as shown in section 4, 
the same evolution is expected even in the RL case. 

Although the reduction of the heat conductivity was originally 
considered due to the electrons scattering by the waves generated by the
RL instability \cite{ram78,hat00}, 
this may not be the case. 
As discussed in above, the wavy magnetic field generated by the instability 
could tend to form the global magnetic field automatically.  
Therefore, the suppression of the heat conductivity
may be determined by the trapping of the electrons by the 
organized magnetic field.
To estimate the suppression of the heat conductivity
quantitatively, we have to know  
what is the final structure of the magnetic field due to 
the self-organization. 
The detail nonlinear studies, numerical simulations for example, 
are desired to answer the questions.

\section{APPLICATION TO THE COLD FRONT IN A3667}

The above results are applied to the cold front found in the cluster of galaxies A3667. 
The temperature and density of the electron changes from  
$k_B T_c=4.1\pm 0.2 \ {\rm keV}$ and   
$n_{e,c}=3.2\pm 0.5 \times
10^{-3} \  {\rm cm^{-3}}$
to $k_BT_h=7.7 \pm 0.8 \ {\rm keV}$  and  
$n_{e,h}=0.82\pm0.12\times 10^{-3} \ {\rm cm^{-3}}$ across the cold
front (Vikhlinin, Markevitch \& Murray 2001a). The width of cold front is $L\sim 5 \ {\rm
kpc}$ at maximum. 
The Coulomb mean free path of the thermal electron in the cold front region is
$\lambda_e=5.4(T/T_{\rm ave})^{2}(n_e/n_{\rm ave})^{-1}\ {\rm kpc}$
where $T_{\rm ave}=(T_h+T_c)/2$ and  $n_{\rm ave}=(n_{e,h}
+n_{e,c})/2$. Thus, $\epsilon\delta_T\sim 1$ in the cold front of A3367.
The growth time scale of the unstable mode in the cold front is 
$\gamma_{\rm max}^{-1}\sim 0.1{\rm sec}$.
Although the exact time scale over which the instability can generate magnetic
fields with the saturation values over the scale of the interfaces is
difficult to estimate, that must be several tens times longer than  the growth time scale.
Therefore, it is expected that the magnetic field is generated almost instantaneously  
compared with any other dynamical time scales in the cluster of galaxies. 
By applying the saturation level discussed in the previous section, 
the saturated magnetic field strength in the cold front should be $B_{{\rm
sp},\perp}\sim \sqrt{\pi}\sqrt{n_e k_BT} \epsilon\delta_T 
\simeq 8 \ (T/T_{\rm ave})^{3/2}(n/n_{\rm ave})^{-1/2} \ \mu {\rm
G}$ where the exact dispersion relation obtained by Ramani and Laval
(1978) was used. 
The obtained value agrees surprisingly  well  with the speculated 
value of  $10 \ \mu {\rm G}$ based on the stability consideration against the KH 
instability. 
The direction of the generated magnetic field should be  almost parallel to the 
cold front surface due to the characteristics of the mode.
As discussed in section 5, the generated wavy magnetic fields could tend to evolve into 
the globally connected field. 
Therefore, we expect that the cold front surface is covered by the globally connected magnetic fields
directed tangential to the surface with the strength of $10 \ \mu {\rm G}$ and
the KH instability is suppressed by these fields. 
Although to suppress the KH instability the existence of  the  magnetic fields
are required only  in the tail of the cold fronts \cite{vik01b},
our model predicts 
the existence of such a strong fields all over  the cold front where the temperature jump exists.
The generated  magnetic field  may significantly reduce the electron mean free path 
in the direction of the temperature gradient and could  suppress the 
heat conduction all over the cold fronts.

\section{DISCUSSION}

We have successfully determined the physical mechanism responsible for one of the two independent
modes of the RL instability,
and shown that the growth mechanism is identical to the Weibel
instability which is well-known as one of the generation mechanism of the 
magnetic field.
Therefore, the RL instability can be also considered as the generation mechanism
of the magnetic
fields in the astronomical situations.
The nonlinear
saturation level of the instability is estimated by considering the
wave-particle interaction.  The evolutions of the magnetic fields 
after the saturation are speculated by referring the previous numerical simulations
which followed the non-linear evolution of the Weibel instability. 
The generated fields 
might be self-organized and evolve into the globally connected
magnetic field.
The results are applied to the cold front in A3667. 
The existence of the magnetic
field of  $\sim 10\mu$G tangential to the surfaces in all over the cold front is predicted. 
The results  surprisingly agree with the predicted nature of the magnetic field
in the cold front 
to explain the smoothness of the cold front surface structure avoiding the KH instability.
Although the suppression of the KH instability requires the existence of the strong magnetic
field only in the tail of the cold front, 
our model predicts   
the existence of such a strong fields all over  the cold front where the temperature jump exists.
The significant suppression of the heat conduction in the cold front is also
expected by the trapping of the electrons by the generated magnetic fields.  
Thus, 
our model predicts that the bow-shaped
structure of the cold fronts should not partially disappear and the
whole structure should be maintained. 

Of course further studies on the RL instability are desired.
The non-linear studies using the numerical simulations specificated to 
the RL instability and the cold front are one of the most important 
remaining subjects.
The identification of the physical mechanism of the mode 1 how the longitudinal 
electric field is generated, is another important remained subject. 

The physics of the RL instability discussed in this paper should be universal 
and the RL instability may play an important role 
in various hot thin plasmas in the universe. 
For example, the temperature inhomogeneities of $\epsilon\delta_T\sim 0.04$
distributed in whole cluster
has been found in almost all clusters of galaxies. 
The RL instability predicts the existence of the cluster halo magnetic 
field with strength of $B\sim 0.04\times\sqrt{n_e k_BT}\sim 0.3\mu$G.
Surprisingly the predicted value just coincides with the 
observationally reported cluster halo magnetic field strength.  
The details of this topic are reported 
in the force coming paper \cite{oka03}.  
We believe that the RL instability may provide 
a new fruitful possibility on the origin of the
cosmic magnetic field.

\section*{Acknowledgments}
The authors gratefully thank T. N. Kato, M. Iijima, M. Takizawa, H. Ohno and Y. Fujita for 
their a lot of fruitful comments, and anonymous referee for
one's constructive comments.

\appendix

\section{Appendix : THE WEIBEL INSTABILITY} 

The Weibel instability is driven by an anisotropic temperature distribution 
and generates the growing 
transverse magnetic standing waves \cite{wei59}. This
instability is well-known as the mechanism of the magnetic fields
generation  from  zero initial magnetic field.
Consider the plasma with an anisotropic temperature such as
\begin{eqnarray}
 f=n_0\frac{1}{\pi^{3/2} v_{{\rm th},\perp}^2 v_{{\rm th},\parallel}} 
\exp\left[- \frac{v_\parallel^2}{v_{{ \rm th}, \parallel}^2 }- \frac{v_\perp^2}{v_{ {\rm th}, \perp}^2 }\right], \nonumber 
\end{eqnarray}
where the subscripts $\parallel, \perp$ respectively denote the parallel
and perpendicular direction to the wave vector. 
The dispersion relation
is obtained by the linear plasma kinetic theory \cite{kra73} as
\begin{eqnarray}
 \omega_r &=& 0, \nonumber \\
 \gamma&=&  \frac{1}{\sqrt{\pi}} k v_{{\rm th,\parallel}} \frac{T_\parallel}{T_\perp} \left[\Bigl(\frac{T_\perp}{T_\parallel}-1\Bigr)- \Bigl(\frac{k c }{\omega_{p}}\Bigr)^2\right]. \nonumber 
\end{eqnarray}
The waves has no real part of the frequency. 
The waves grow when $T_\perp > T_\parallel$.
The excited waves are transverse magnetic waves with no electric field. 

The physical mechanism of the Weibel instability is first identified  by  
Fried (1959).    
Consider the simple situations  as shown in Fig. 4. 
The electrons and their initial velocities are expressed as the
filled circles and the dotted arrows.
In the case A, 
the simple anisotropic initial electron distribution function
\begin{eqnarray*}
f_0(\vec{v})=n_0u\delta(v_\parallel)\delta(v_\perp^2-u^2)\delta(v_z)
\end{eqnarray*}
is assumed where $n_0$ is the electron number density. 
This   represents an extreme case of  $T_\perp
 \neq 0, T_\parallel = 0$. 
In the case B, 
the  anisotropic initial electron distribution function
\begin{eqnarray*}
f_0(\vec{v})=n_0w\delta(v_\parallel^2-w^2)\delta(v_\perp)\delta(v_z)
\end{eqnarray*}
is assumed.
This   represents an extreme case of  $T_\perp
 = 0, T_\parallel \neq 0$. 
The directions of the perturbed magnetic
fields are perpendicular to the paper : $\odot$ \&
$\otimes$ represent that the fields project from the paper and rush in the paper, respectively.
The size of the circle is proportional to the field strength at each position.

Firstly, the qualitative explanation of the physical mechanism is reviewed.
The electrons' 
orbits  are deflected due to the introduction of the perturbed  magnetic field 
as shown by the dashed arrows. 
The amount of the deflection is proportional to the
field strengths at the each position. 
We first discuss the Case A.  Consider the electrons  (a) and  
(a') which initially have the negative
velocity in the $x_\perp$ direction. They respectively carry the positive 
electric 
currents  in the $x_\perp$ direction out from and into the shaded region.
Because of the difference of
the field strengths, the net positive current in the $x_\perp$
direction is carried out from the region in a certain infinitesimal time 
interval. 
In other words the net negative electric current
is carried  into the region by the perturbed motion of these electrons. 
The electrons  (b) and (b')  also carry the net negative 
electric current into the shaded region. 
Therefore, the net negative electric current is developed
in the shaded region due to the injection of the perturbed 
magnetic field.  
Since the values of the induced electric
currents are larger where the gradient of the 
magnetic field strength is larger, 
the current fields illustrated in  Fig. 4 are set up where 
the length of the arrows are proportional to the amplitudes of the currents.
The magnetic
field is produced around each electric current according to the Amp\`ere's law as illustrated in Fig.4.
The growth of the  magnetic field is determined by the superposition of these 
fields. 
For example, at point $x_1$ the net magnetic field 
generated by the current fields is  normal to the paper toward 
readers
and amplifies the injected perturbed field. 
Since the excited fields amplify the injected perturbed
fields everywhere, the perturbed field grows in the Case A. 
On the other hand, the net induced electric currents
direct  opposite direction to Case A in the Case B. 
Therefore, the perturbed magnetic field is damped in the Case B.
Hence, when $u>w$ ($T_\perp
>T_\parallel$) the magnetic field perturbation  grows and when $u<w$ ($T_\perp
<T_\parallel$) the magnetic field perturbation is damped.
While this simple explanation shows qualitatively how the growth occurs, 
the role of the electromagnetic induction for the growth of the 
instability, which was neglected in the above discussion,
should be examined since it acts to reduce the growth of the magnetic field.  

Next, the semi quantitative explanation of the instability is given to see 
the role of the induction.  For this purpose, only the case A is considered.
Suppose that an initial perturbed magnetic field is 
\begin{eqnarray*}
B_z&=&Be^{ikx_\parallel}.
\end{eqnarray*}
The $x_\parallel$ and $x_\perp$ components 
of the equation of motion of the electron  are 
given by 
\begin{eqnarray*}
m_e{d v_\parallel\over dt}&=&-e {v_\perp\over c}B_z,\\
m_e{d v_\perp\over dt}&=&e {v_\parallel\over c}B_z-{ie\over k c}{\partial B_z\over 
\partial t},
\end{eqnarray*}
respectively, where the last term in $x_\perp$ component represents
the force due to the inductive electric field. 
Therefore, the time derivative of the electric current 
flux is given by 
\begin{eqnarray*}
{\partial <j_\perp v_\parallel>\over \partial t}
&=&e^2 {u^2\over c m_e}B_z,
\end{eqnarray*}
where $j_\perp=-ev_\perp$ is the $x_\perp$ component of the 
electric current carried by an single electron and 
$<X>={1\over n_0}\int d^3v X(v) f(v)$ is the average over the velocity space.
This current flux and the inductive electric field cause a change in the 
mean value of $j_\perp$ as
\begin{eqnarray*}
{\partial <j_\perp>\over \partial t}&=&-{\partial <j_\perp v_\parallel>\over \partial x_\parallel}
+e^2 i{1\over m_ek c}{\partial B_z\over \partial t}.
\end{eqnarray*}
The Amp\`ere's law provides 
\begin{eqnarray*}
n_0<j_\perp>&=&-{ick\over 4\pi}B_z,
\end{eqnarray*}
where the displacement current is neglected since the low frequency mode is only 
considered. 
By combining these equations, we obtain
\begin{eqnarray*}
\left[1+{\omega_{p}^2\over c^2k^2}\right]{\partial^2 B_z\over \partial t^2}
&=&{\omega_{p}^2\over c^2}u^2B_z.
\end{eqnarray*}
The dispersion relation is obtained as 
\begin{eqnarray*}
\gamma&=&{(\omega_{p}/c) u\over \sqrt{1+{\omega_{p}^2\over c^2k^2}}}.
\end{eqnarray*}
It shows that the perturbation can grow. 
The second term in the $[\;]$ of the left hand side is 
coming from the induction term.
Therefore, the electromagnetic induction actually plays a role of the 
inertia for the growth of the perturbation and reduces the growth rate,
but never stop the growth of the perturbation. 
The obtained growth rate is exactly the same as the result
deduced by the full linear analysis based on the plasma kinetic theory
where the induction effect is also taken into account (e.g. Melrose 1986).

\clearpage

\begin{figure}
\begin{center}
\includegraphics[scale=1.6]{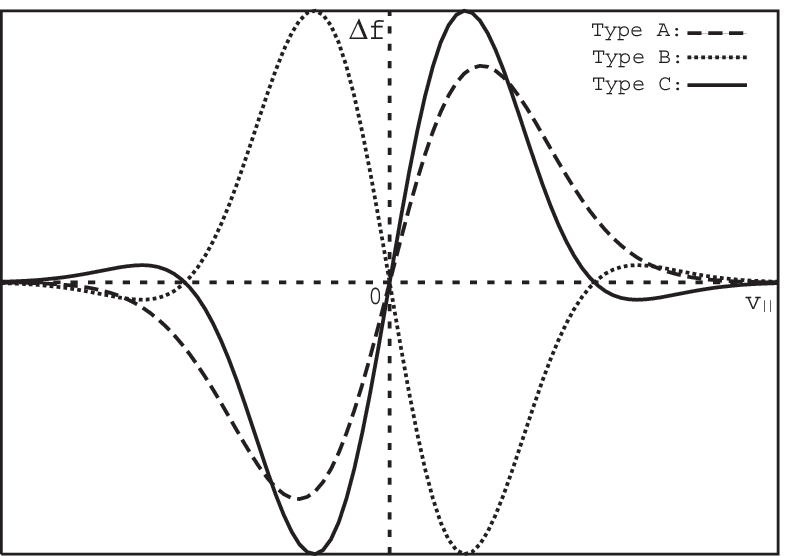}
\caption{The possible candidates as the deviated part of the velocity
 distribution function from the Maxwell-Boltzmann.}
\end{center}
\end{figure}

\clearpage

\begin{figure}
\begin{center}
\includegraphics[scale=1.6]{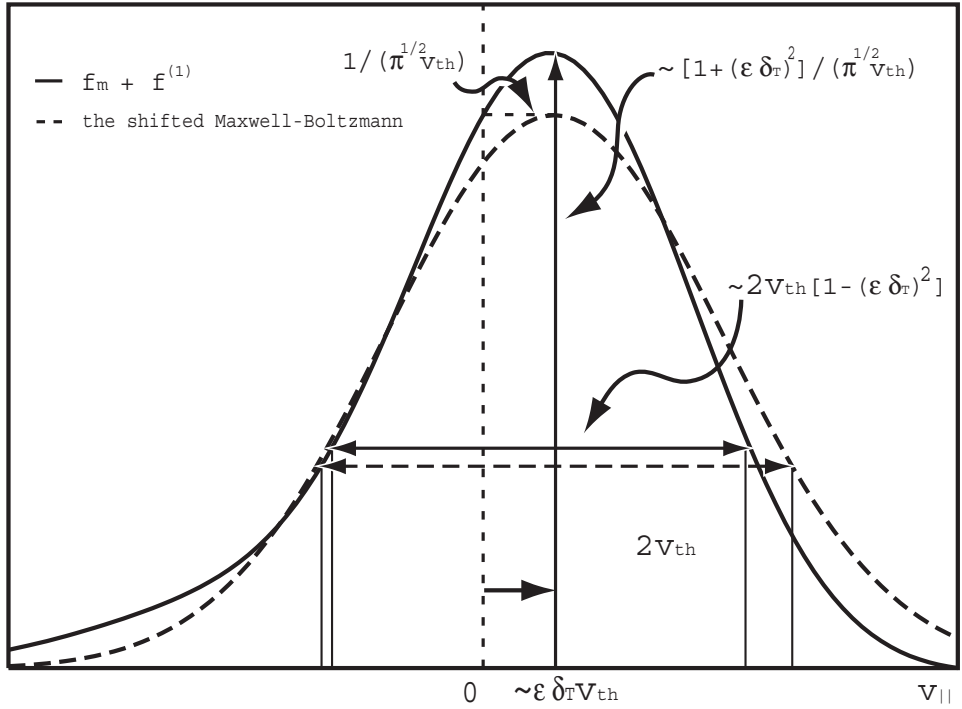}
\caption{The solid line is the $v_\parallel$ section of the total velocity distribution function $f_m+
f^{(1)}$ in the plasma with
the temperature gradient. The peak position is  shifted by $\sim \epsilon \delta_T v_{{\rm th}}$ from the
Maxwell-Boltzmann. The peak value is increased by $\sim 1+
\epsilon^2 \delta_T^2$ compared with the  Maxwell-Boltzmann. For
comparison, the Maxwell-Boltzmann velocity distribution function shifted
by $\sim \epsilon \delta_T v_{{\rm th}}$ drawn in the dashed line. The
velocity distribution function gets thinner in $v_\parallel$
direction. This can be interpreted as the decrease of the effective
temperature by  $\sim 1-\epsilon^2 \delta_T^2$ in the direction of the
temperature gradient.}
\end{center}
\end{figure}

\clearpage

\begin{figure}
\begin{center}
\includegraphics[scale=1.6]{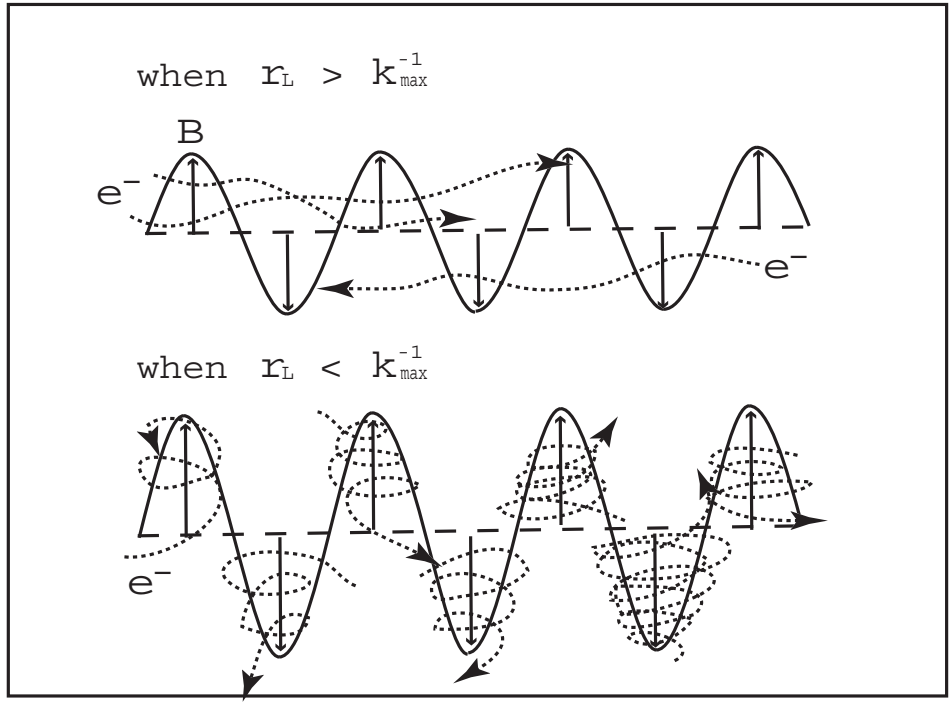}
\caption{The nonlinear saturation by the wave-particle interaction. 
The top panel: When $r_L > k_{{\rm max}}^{-1}$, the thermal electrons
travel throughout the waves but their orbits are randomly disturbed by the wavy
magnetic fields.  The bottom panel: Once $r_L$ becomes  smaller than
$k_{{\rm max}}^{-1}$, the thermal electrons are
trapped by the fields and feels net non zero fields. }
\end{center}
\end{figure}

\begin{figure}
\begin{center}
\includegraphics[scale=1.6]{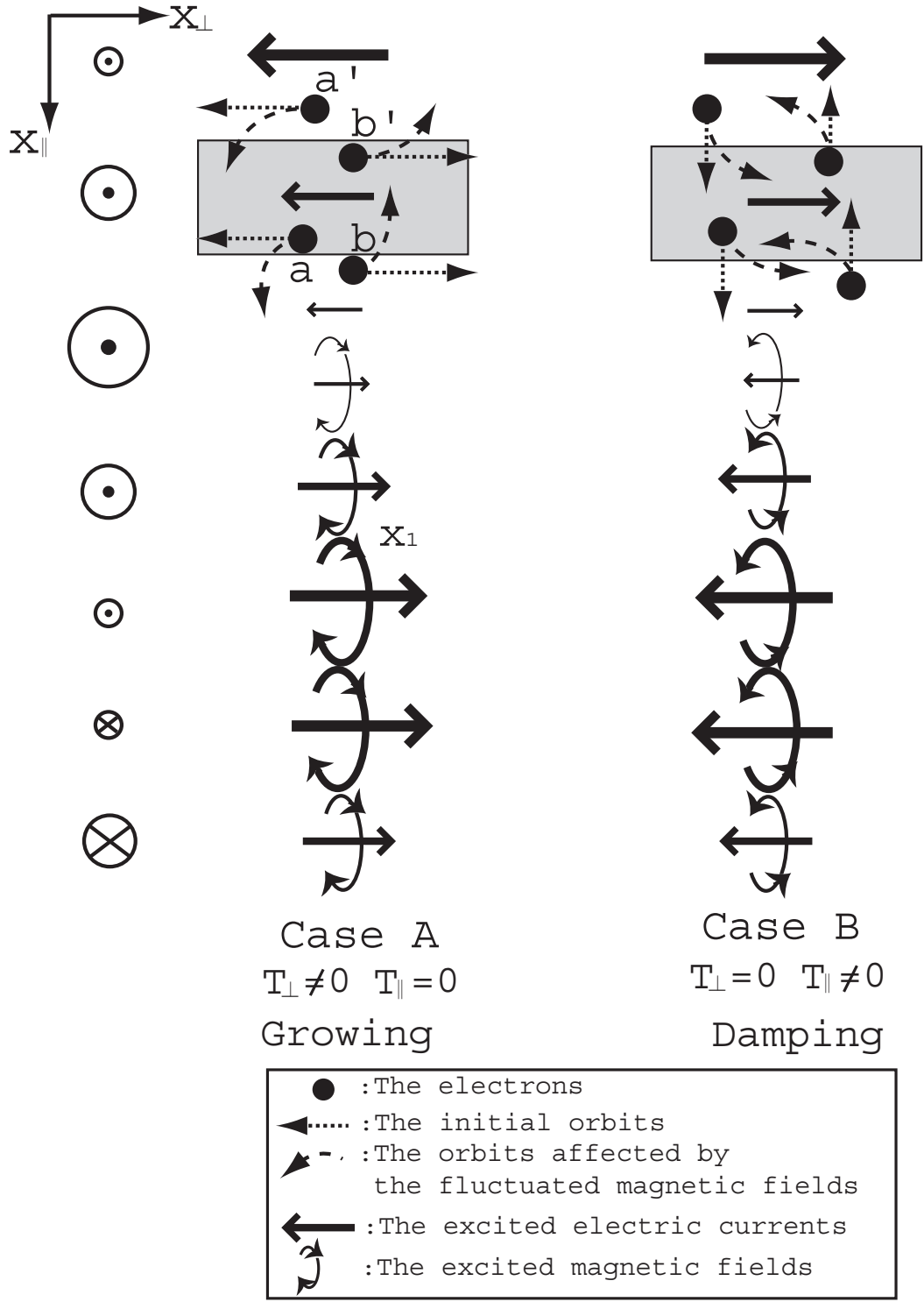}
\caption{The physical mechanism of the Weibel instability}
\end{center}
\end{figure}

\end{document}